\documentstyle[pra,aps,amsmath,graphicx]{revtex}
\begin{document}
\draft
\twocolumn[\hsize\textwidth\columnwidth\hsize\csname 
@twocolumnfalse\endcsname
\title{Quantum erasure by transverse indistinguishability} 
\author{P. H. Souto Ribeiro$^{1}$ \cite{ca}, S. P\'adua$^{2}$, 
and C. H. Monken$^{2}$}
\address{$^{1}$ Instituto de F\'{\i}sica, 
Universidade Federal do Rio de 
Janeiro, Caixa Postal 68528, Rio de Janeiro, RJ 22945-970, Brazil\\
$^{2}$ Departamento de F\'{\i}sica, 
Universidade Federal de Minas Gerais, Caixa Postal 702, 
Belo Horizonte, MG 30123-970, Brazil} 
\date{\today}
\maketitle
\begin{abstract}
We show that the first experiment with double-slits and twin photons
detected in coincidence can be understood as a quantum eraser.  The
``which path'' information is erased by transverse
indistinguishability obtained by means of mode filtering in the twin
conjugated beam.  A delayed choice quantum eraser based on the same
scheme is proposed.
\end{abstract}
]
\section{INTRODUCTION}

Studying Quantum Physics experimentally has become much easier with
the use of the twin photons produced in the
spontaneous parametric down-conversion process.
Many of the experiments performed so far, are realizations
of different kinds of interferometers.  These interferometers allow us
to improve our understanding of the nature of light and matter.  It is
convenient to think these interferometers, as divided in two
categories: the ones utilizing the longitudinal degrees of freedom of
the field and the ones utilizing its transverse degrees of freedom.

We call longitudinal interferometers, the ones of the type of
Michelson's and Mach-Zehnder's interferometers.  A large
number of experiments with twin photons has been performed with them. 
Some examples are those of Refs.  \cite{1,2,3}.  The transverse
interferometers, are the ones of the kind of the double-slit or Young
interferometers.  Some examples are given in Refs. 
\cite{4,5,6,7,8,9,10}.  These interferometers are important
because they are perfect realizations of some {\em
gedankenexperiments}, used in the discussion of important issues on
the foundations of Quantum Mechanics.  The transverse interferometers
are very robust, quite stable and phase differences can be controlled
by simple displacement of detectors in most of the cases.  For this
reason, they have become an important tool in the field of
multiparticle interferometry.

In this paper, we analyze a transverse interferometer utilizing the
twin photons of the down-conversion.  As far as we are concerned, it
was the first interferometer with twin photons and a
double-slit\cite{5}.  The main result of that work, was to show that
interference fringes were observed in the coincidence counting rate,
while at the same time, intensity interference fringes could not be
observed.  Our aim here is to revisit this experiment and show that it
can be viewed as a quantum eraser\cite{11}.  Even though other
experiments performed in the past could also be interpreted in the
same way as a quantum eraser, this experiment was probably one of the
first utilizing twin photons.  The fact that twin photons were
employed stresses the quantum character of the {\em which path}
information erasure, since it is performed by quantum state projection
of a nonlocal wave function.

\section{THE DOUBLE-SLIT EXPERIMENT WITH TWIN PHOTONS}

Let us briefly recall the main idea behind the experiment described in
Ref.  \cite{5}.  Fig.  \ref{fig1} shows a sketch of the set-up.  Twin
photons are produced in the process of spontaneous parametric
down-conversion and they are detected in coincidence.  The signal
photon is passed through a double-slit and the idler photon goes
straight to the detector.  Coincidence fringes are detected by displacing
transversely the signal detector, which is after the slits and keeping the idler
detector fixed.  It was shown that the coincidence profile exhibited
interference fringes, even when intensity fringes were not observed.

\begin{figure}[h]
\vspace*{4.5cm}
\hspace*{1cm}
%\special{eps:qe1.eps x=5cm y=5cm}
\caption{Two-slit interference experiment with twin photons.}
\label{fig1}
\end{figure}

As it is known, and it was demonstrated for the parametric
down-conversion\cite{4}, the second order coherence, which defines the
visibility of the intensity fringes, is dependent on the geometrical
properties of the light source.  If we have a small source,
approximately spherical waves are generated and if the slits are far
enough, the second order coherence is almost perfect.  Then, intensity
interference fringes are obtained in a double-slit experiment with
unity visibility.  If we have an extended source instead, the degree
of coherence can be smaller than one, as well as the visibility of the
intensity interference fringes.  The degree of coherence depends
strongly on the source length and the visibility of the fringes
depends on the separation between slits and the distance between
source and slits.

However, if twin photons are used and coincidence detection is
performed, the coherence function is now a fourth order one. 
The degree of coherence becomes dependent on the way the idler beam is
detected.  The coincidence interference fringes also become dependent
on it.  In particular, it was shown\cite{5} that the detection through
a small idler aperture leads to an improvement on the coincidence fringes
visibility and a large idler aperture tends to destroy the interference
fringes.  Even if we do not have second order interference, fourth
order coherence can be obtained by the use of a small area detector
for the idler beam.  This is the principle of a quantum eraser.

\section{THE QUANTUM ERASER ALGORITHM}

For the sake of simplicity, we will demonstrate the quantum erasure
aspect of the above mentioned experiment by comparing it with a
simpler system.  We will introduce a quantum erasure algorithm.  Any
experience following this procedure, can be understood as a quantum
erasure.  The analysis is the same as in reference  \cite{12}.

\begin{figure}[h]
\vspace*{4.5cm}
\hspace*{1cm}
%\special{eps:qe2.eps x=5cm y=5cm}
\caption{Quantum eraser algorithm.}
\label{fig2}
\end{figure}

The quantum eraser algorithm has three steps:

\begin{enumerate}
\item
A superposition of state is produced and a phase dependent measurement
is performed on the state.  See Fig. \ref{fig2}a for example.  A small
light source emits a photon.  The photon passes through a double-slit. 
The slits are labeled 1 and 2 and the state of the photon after the
slits is given by:

\begin{equation}
\label{eq1}
|\Psi\rangle = \frac{1}{\sqrt{2}} ( | \psi_{1} \rangle + | \psi_{2}
\rangle ).
\end{equation}

By detecting the photon after the slits, in different positions in a
plane transverse to the propagation direction, a phase dependent
measurement is performed.  The phase difference between states $|
\psi_{1} \rangle $ and $| \psi_{2} \rangle $ is proportional to the
path difference from each slit to the detector.  Interference
fringes are then observed.
\item
The second step is to provide a possibility of having {\em which path}
information.  In Fig.  \ref{fig2}b, this is performed by placing wave
plates in front of each slit, so that if the photon passes through
slit 1, it will be $L$ (left) circularly polarized and if it passes
through slit 2, it will emerge $R$ (right) circularly polarized.  As
it is known, the mere possibility of having {\em which path}
information destroys the interference pattern.  In terms of the state
of the system, states $| \psi_{1} \rangle $ and $| \psi_{2} \rangle$ have been
entangled with an internal degree of freedom (the polarization), which
is used to identify the path:

\begin{equation}
\label{eq2}
|\Psi\rangle = \frac{1}{\sqrt{2}} ( | \psi_{1} \rangle|L\rangle  + | 
\psi_{2} \rangle|R\rangle ).
\end{equation}

Notice that a polarization independent detection leads to an incoherent sum of
the intensities due to each polarization. This leads to no interference.
\item
Finally, the third and last step consists of erasing the {\em which
path} information.  In Fig.  \ref{fig2}c, passing the photon through
a linear polarizer performs the erasure.  This is equivalent to a
projection of the system onto a state which is not entangled with the
polarization anymore:

\begin{equation}
\label{eq22}
|\Psi\rangle = \frac{1}{\sqrt{2}} ( | \psi_{1} \rangle + | \psi_{2} 
\rangle)|\theta\rangle,
\end{equation}
where $\theta $ represents the angle of the polarizer.  The
interference fringes are recovered.
\end{enumerate}

The above state can be understood as the state of the system just
before detection.  In fact, it is the passage of the photon through
the polarizer, which projects its state onto a linear polarization
state.

In next section, we will show that the experiment of reference \cite{5}
follows the above algorithm.

\section{QUANTUM ERASURE BY TRANSVERSE INDISTINGUISHABILITY}

A sequence of steps associated with the experiment of Fig.  \ref{fig1}
is now shown in Fig. \ref{fig3}.  In Fig. \ref{fig3}$a$, we focus on the
signal beam where the double-slit plane is far enough from the down-conversion crystal, 
so that it emits approximately like a point source.  From the detector's
point of view, a photon that is detected at a particular point $P$ on
the detection plane has a probability amplitude $\psi_{1}(P,P')$
associated with its passage through slit 1, assuming that it was
generated at point $P'$ on the source plane.  Analogously,
$\psi_{2}(P,P')$ is the amplitude associated with generation at point
$P'$, passage through slit 2 and detection at point $P$.  For a fixed
detection point $P$, $\psi_{1}(P,P')$ and $\psi_{2}(P,P')$ are the
diffraction amplitudes of the slits 1 and 2, respectively, on the
source plane, as if they were illuminated by a point source located at
$P$.  For the configuration just described, since the source is far
away from the slits, the overlap between $\psi_{1}(P,P')$ and
$\psi_{2}(P,P')$ is much larger than the source dimensions.  So,
passage through slit 1 or slit 2 are indistinguishable
possibilities and interference arises.  The state of the signal
photon after the slits is given by:

\begin{equation}
\label{eq4}
|\psi\rangle_{s} = \frac{1}{\sqrt{2}} ( | \psi_{1} \rangle_{s} + | 
\psi_{2} \rangle_{s} ).
\end{equation}
The first step is accomplished.

\begin{figure}[h]
\vspace*{4.5cm}
\hspace*{1cm}
%\special{eps:qe3.eps x=5cm y=5cm}
\caption{Quantum erasure by transverse indistinguishability.}
\label{fig3}
\end{figure}

Now, if we move the slits screen close to the crystal, $\psi_{1}(P,P')$ 
and $\psi_{2}(P,P')$ do not overlap anymore (Fig. \ref{fig3}$b$). That 
is, passage through slit 1 (2) implies that the photon was generated 
in region $A$ ($B$) of the source. Since the interaction volume in 
spontaneous parametric down-conversion behaves like an incoherent 
source \cite{4}, emission from regions $A$ and $B$ are, in principle, 
distinguishable events. No interference is observed in this 
configuration.  The state of the photon after the slits is now labeled by
the emitting regions A and B:

\begin{equation}
\label{eq4a}
|\psi\rangle_{s} = \frac{1}{\sqrt{2}} ( | \psi_{1}\rangle_{A,s} + | 
\psi_{2} \rangle_{B,s} ).
\end{equation}

We are going to make use of the idler photon, which was not shown in
Fig.  \ref{fig3}a, to obtain the {\em which path} information.  It was
shown by Klyshko \cite{klyshko} that signal and idler photons in
spontaneous parametric down-conversion have a high spatial correlation
on the source, that is, they are generated in the same point in the
interaction region.  The idler photon will be labeled by $A$($B$) if it is
emitted at the region $A$($B$) of the crystal.  Then the state of the twin
photons is given by:

\begin{equation}
\label{eq5}
|\Psi\rangle = \frac{1}{\sqrt{2}} ( | \psi_{1} \rangle_{s} | A \rangle_{i} +
| \psi_{2} \rangle_{s} | B \rangle_{i} ),
\end{equation}
where the index $s$ stands for the signal beam and $i$ for the idler. 
Step two is accomplished.  Note that now, the system is entangled with
an auxiliary system, the idler photon, which serves as a {\em label}.

The field generated by an incoherent source can be viewed as a
statistical ensemble of coherent fields \cite{mandelwolf}.  It is
known from diffraction theory that in a transverse plane located in
the far zone, the spatial distribution of each realization of this
ensemble is proportional to the Fourier transform of its distribution
on the source plane.  By inverting the Fourier transform, for example
with the help of a lens, an image of the source is formed in the far
zone, and it is possible to detect which zone ($A$ or $B$) of the
source the idler photon was emitted from.

Let us now see how this path information can be erased.  If a small
aperture is placed on the idler detection plane, it will act as a
spatial filter.  In other words, it will band-limit the Fourier
transform of each coherent realization of the ensemble.  The smaller
the aperture, the narrower the spatial bandwidth of the filtered
field.  Since the details of an image are carried by the high
frequency components of its angular spectrum \cite{mandelwolf}, the
strong spatial filtering produced by a small enough aperture will make
impossible to retrieve spatial information about the source.  Then,
this spatial filtering erases {\em which path} information from the
point of view of the idler detector.

Expanding states $| A \rangle_{i}$ and $| B
\rangle_{i}$ in terms
of their transverse Fourier components we have

\begin{eqnarray}
\label{eq6}
 | A \rangle_{i} = \int {\mathcal A} ( \boldsymbol{q} )\ 
 |\boldsymbol{q} \rangle_{i} \ d \boldsymbol{q} \, ; \\ \nonumber 
 | B \rangle_{i} = \int {\mathcal B} ( \boldsymbol{q} )\ 
 |\boldsymbol{q} \rangle_{i} \ d \boldsymbol{q}.
\end{eqnarray}

Detection of the idler photon through a small pinhole corresponds to
a projection of the states $| A \rangle_{i}$ and $| B \rangle_{i}$
onto approximately the same state $| \boldsymbol{q}_{o} \rangle$. 
$\boldsymbol{q}_{o}$ represents a given transverse wave vector
component \cite{mandelwolf}. In the limit of a point detector,
it is exactly the same state vector. Then, the state of the twin
pair becomes:

\begin{equation}
\label{eq7}
|\Psi\rangle = \frac{1}{\sqrt{2}} (| \psi_{1} \rangle_{s} 
+|\psi_{2}\rangle_{s})|\boldsymbol{q}_{o}\rangle_{i}.
\end{equation}

Interference is retrieved by performing coincident detection of signal
and idler photons (Fig.  \ref{fig3}$c$).  Step three is accomplished.

Again, the above state describes the photon pair just before
detection, and it is the passage of the idler photon through the
pinhole which projects the system onto that state.  When a large
aperture is used for the detection of the idler photon, the
interference disappears again.  The larger the detection area is, the
larger is the number of Fourier components that are taken.  In the
limit where all Fourier components are detected, states $| A
\rangle_{i}$ and $| B \rangle_{i}$ can be completely reconstructed.
The state of the system is again given by expression (\ref{eq5})
and the idler photon again serves as a label.

In the usual discussions about the quantum eraser, the state of the
which-path detector is described in a two-dimensional Hilbert space. 
To accomplish erasure, that state is projected onto a superposition of
two ``pointer'' states, such that  a symmetric superposition gives rise to
``fringes'', whereas a antisymmetric one gives rise to 
``antifringes'' \cite{11}. 
In the context discussed here, the state of the idler photon (the
which-path detector) is described by a continuum of modes (its 
Fourier decomposition). Instead of ``fringes'' and ``antifringes'', 
we have a continuum of interference patterns labeled by the 
selected transverse Fourier component $\boldsymbol{q}_{o}$. 

\section{A DELAYED CHOICE QUANTUM ERASER SCHEME}

\begin{figure}[h]
\vspace*{3cm}
%\hspace*{.5cm}
%special{eps:qe4.eps x=7cm y=3cm}
\caption{Delayed choice quantum eraser.}
\label{fig4}
\end{figure}

The experimental set-up described in Fig.  \ref{fig1} can be easily
changed, in order to obtain a delayed choice quantum eraser.  See Fig. 
\ref{fig4}.  By introducing a beam splitter in the path of the
conjugated beam and sending each half to two different detectors, it
is possible to choose between interference and non-interference.  One
of the detectors has a small aperture and the other a large aperture. 
These detectors and beam splitters can be put very far from the crystal,
so that the signal photon passes through the double-slit before the idler photon
gets to the beam splitter.  A photon count in the large aperture
detector, implies in no interference and a photon count in the small
aperture detector implies in interference.  The decision between
interference and no interference is made after the photon has passed
through the slits.

In the original experiment, the detection through large and small
apertures is not performed simultaneously.  However, there are no
reasons for the results to be different.  In fact, we note that in
this scheme, the decision of interference or not is not made in the
beam splitter, but in the detector.  Finally, it is worth mentioning
that our double-slit quantum eraser satisfies all the criteria for a
true quantum eraser as stated by Kwiat, Steinberg and Chiao
\cite{kwiat}: a delayed choice scheme can be implemented, it employs
single particles and the distinguishing information is carried
separately from the interfering particle.
 
\section{CONCLUSION}

We have analyzed a double-slit interference experiment from a point of
view of a quantum eraser.  We show that the measurements performed,
followed a certain quantum eraser algorithm.  We also show that a
delayed choice quantum eraser could be easily obtained by the
insertion of a beam splitter on the idler beam and detection with
different detection areas.  This experiment, was probably the first
realization of a quantum eraser dealing with the transverse degrees of
freedom of the field.

\section{ACKNOWLEDGMENT}
The authors acknowledge financial support from the Brazilian 
agencies CNPq, PRONEX, FAPERJ, FUJB and FAPEMIG. 
%\bibliographystyle{prsty}
%\bibliography{artqi}

%\begin{figure}[h]
%\includegraphics[width=7cm]{qe1.eps}
%\caption{Two-slit interference experiment with twin photons.}
%\label{fig1}
%\end{figure}

%\begin{figure}[h]
%\includegraphics[width=7cm]{qe2.eps}
%\caption{Quantum eraser algorithm.}
%\label{fig2}
%\end{figure}

%\begin{figure}[h]
%\includegraphics[width=7cm]{qe3.eps}
%\caption{Quantum erasure by transverse indistinguishability.}
%\label{fig3}
%\end{figure}

%\begin{figure}[h]
%\includegraphics[width=7cm]{qe4.eps}
%\caption{Delayed choice quantum eraser.}
%\label{fig4}
%\end{figure}

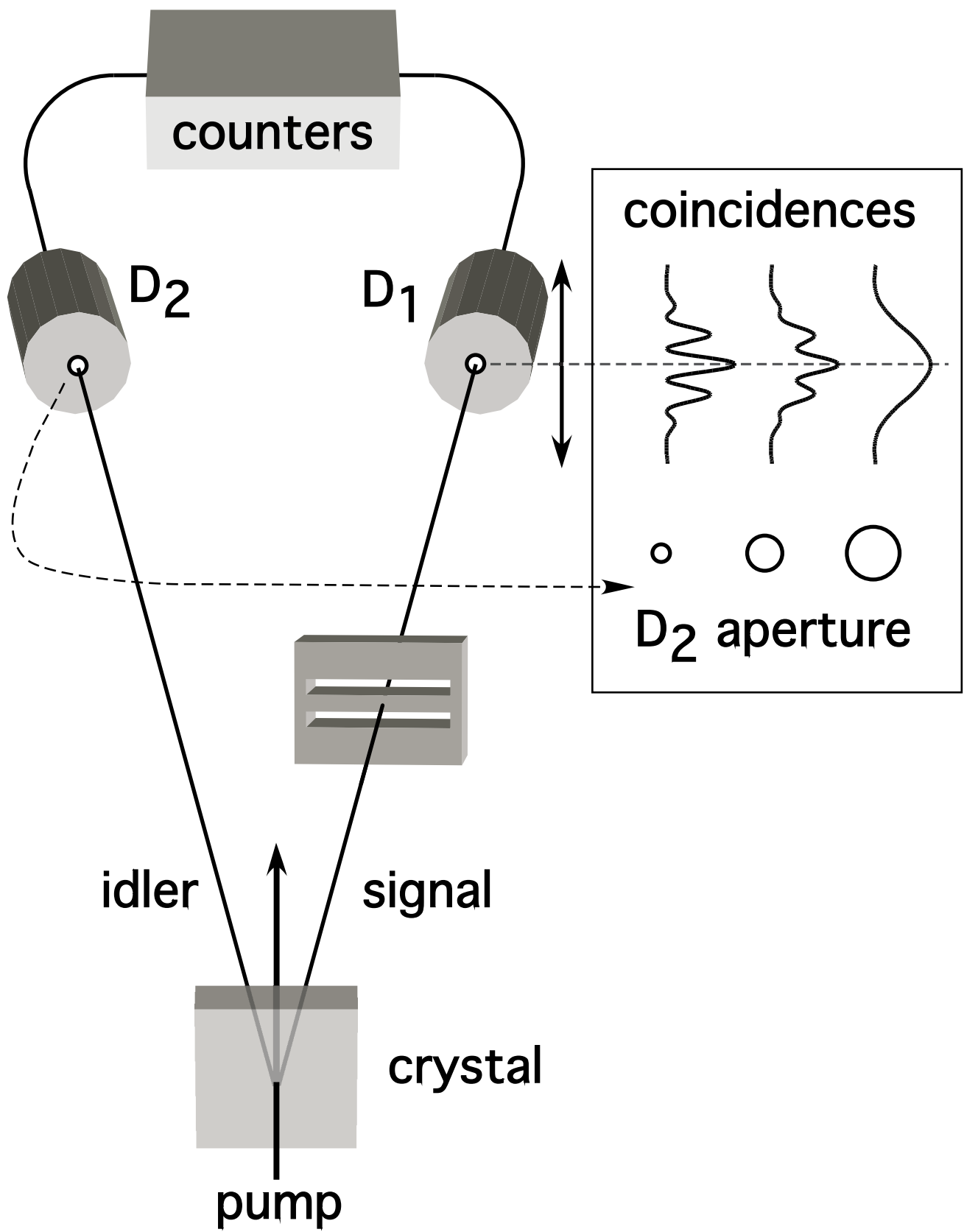
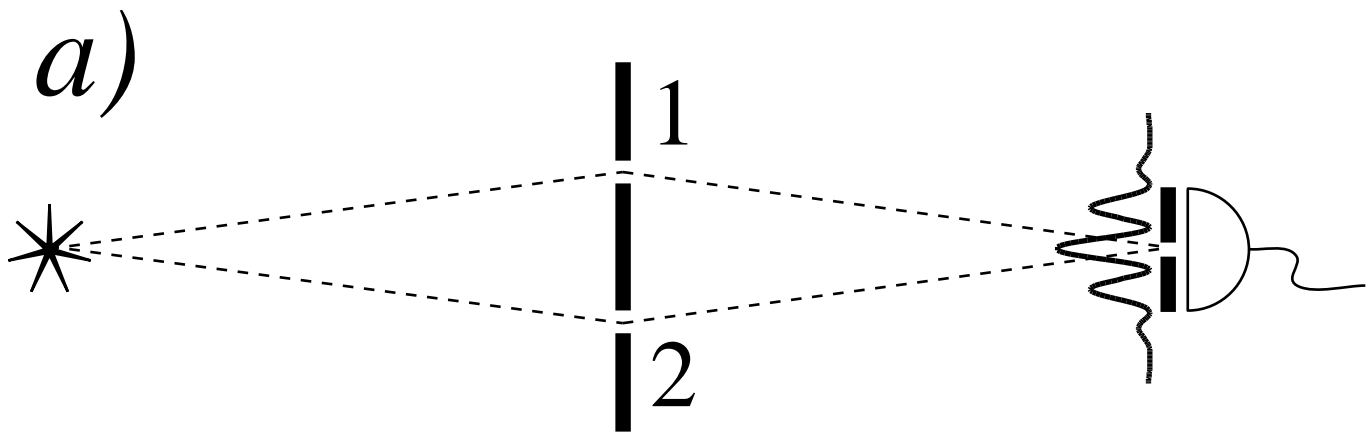
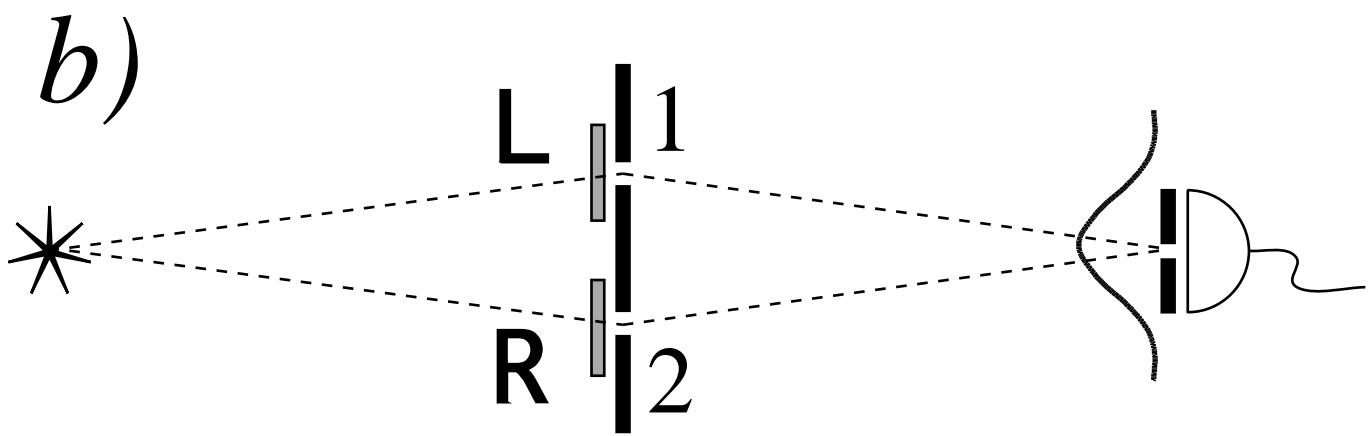
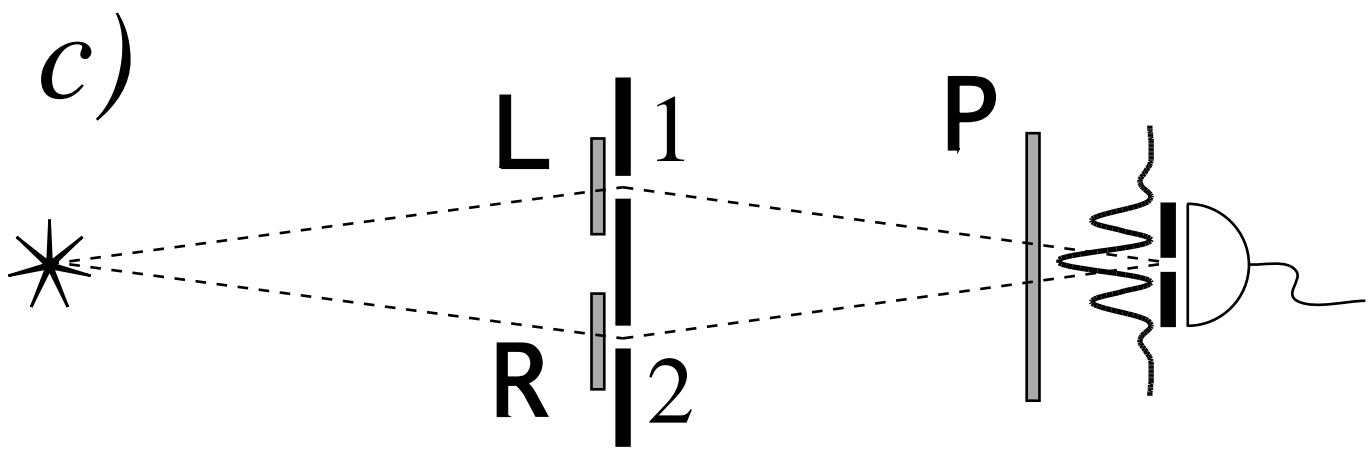
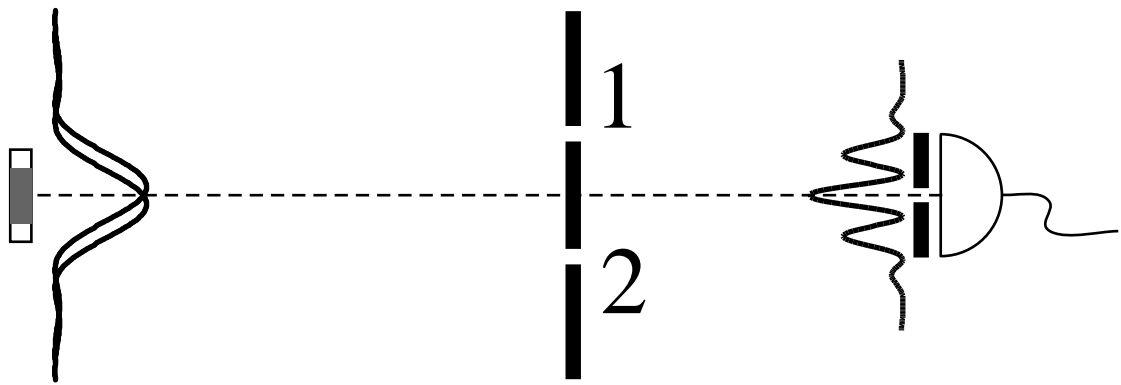
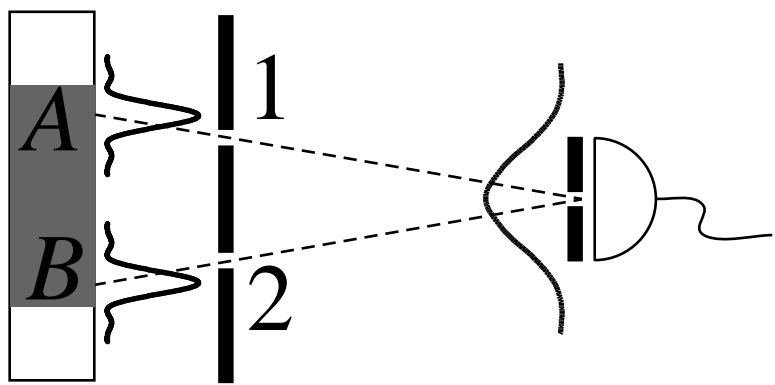
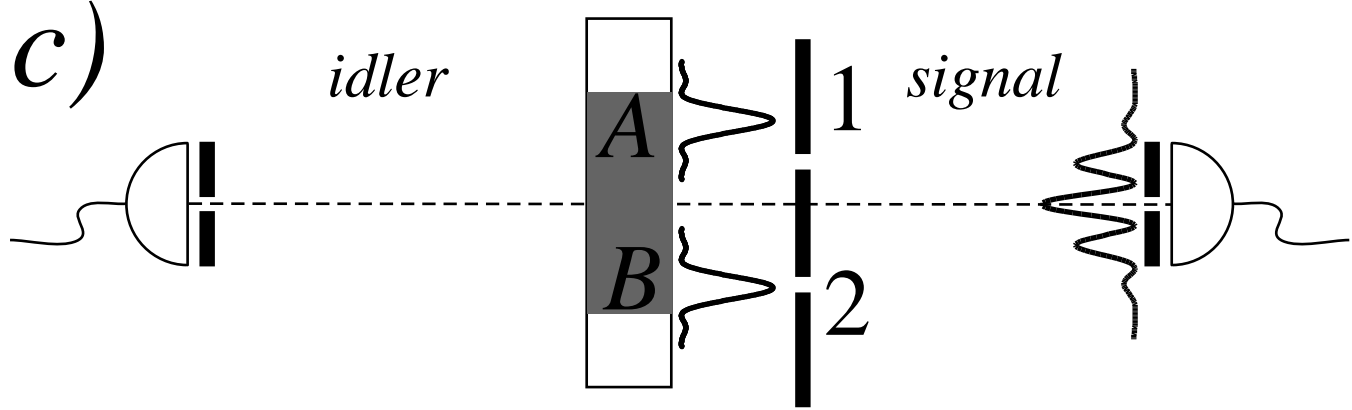
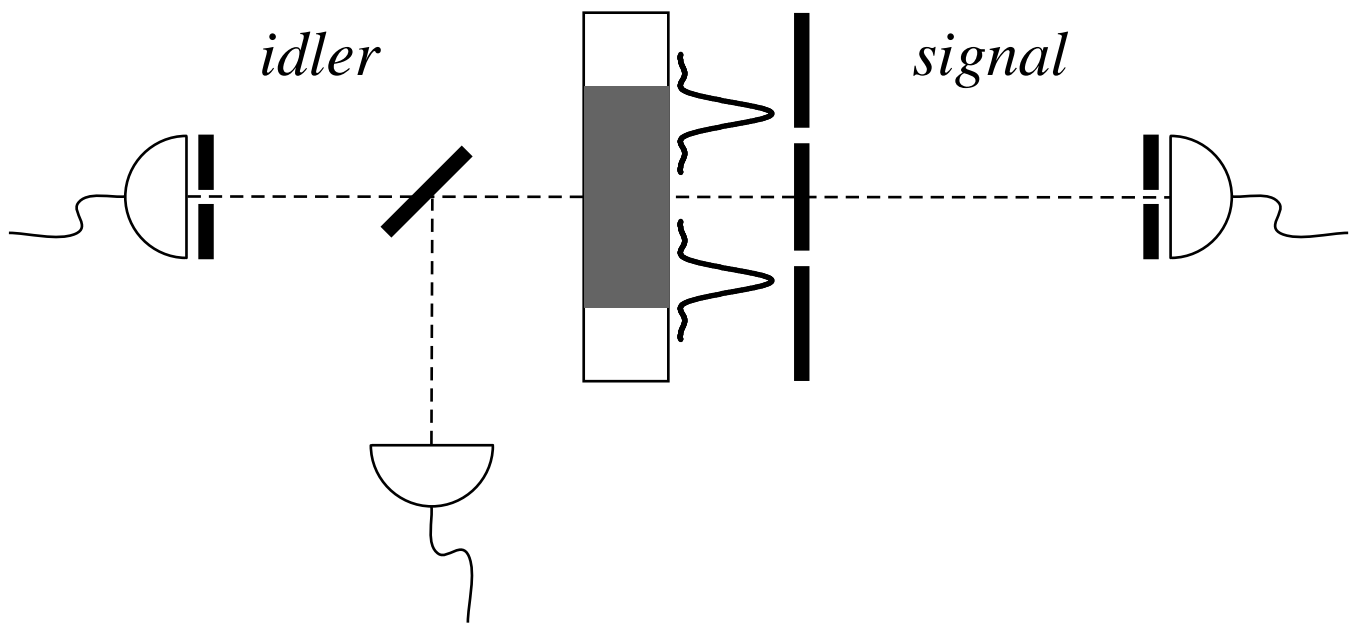
\end{document}